\documentclass[prl,twocolumn,superscriptaddress,showpacs,floatfix]{revtex4}
\usepackage{graphicx,amsmath,amssymb,bm}
\usepackage{amsfonts}

\begin{document}

\title{Continuum effects and three-nucleon forces in neutron-rich oxygen isotopes}

\author{G.~Hagen}
\affiliation{Physics Division, Oak Ridge National Laboratory,
Oak Ridge, TN 37831, USA}
\affiliation{Department of Physics and Astronomy, University of
Tennessee, Knoxville, TN 37996, USA}
\author{M.~Hjorth-Jensen}
\affiliation{Department of Physics and Center of Mathematics for Applications, University of Oslo, N-0316 Oslo, Norway}
\affiliation{National Superconducting Cyclotron Laboratory, Michigan
  State University, East Lansing, MI 48824-1321, USA}
\affiliation{Department of Physics and Astronomy, Michigan State University, East Lansing, MI 48824, USA}
\author{G.~R.~Jansen}
\affiliation{Department of Physics and Center of Mathematics for Applications, University of Oslo, N-0316 Oslo, Norway}
\author{R.~Machleidt}
\affiliation{Department of Physics, University of Idaho, Moscow, ID 83844, USA}
\author{T.~Papenbrock}
\affiliation{Department of Physics and Astronomy, University of
Tennessee, Knoxville, TN 37996, USA}
\affiliation{Physics Division, Oak Ridge National Laboratory,
Oak Ridge, TN 37831, USA}

\begin{abstract}
  We employ interactions from chiral effective field theory and
  compute binding energies, excited states, and radii for isotopes
  of oxygen with the coupled-cluster method. Our calculation
  includes the effects of three-nucleon forces and of the particle
  continuum, both of which are important for the description of
  neutron-rich isotopes in the vicinity of the nucleus $^{24}$O. Our
  main results are the placement of the neutron drip-line at $^{24}$O,
  the assignment of spins, parities and resonance widths for
  several low-lying states of the drip-line nucleus, and an efficient
  approximation that incorporates the effects of three-body
  interactions. 
\end{abstract}


\maketitle

{\it Introduction.} -- Neutron-rich oxygen isotopes are particularly
interesting nuclei.  First, the nuclei $^{22}$O and $^{24}$O exhibit
double magicity at the neutron numbers $N=14$ and $N=16$,
respectively, see for example
Refs.~\cite{thirolf2000,hoffman2009a,kanungo2009}.  Second, oxygen is
the heaviest element for which the neutron drip line is established
experimentally.  The recent
experiments~\cite{hoffman2009b,lundeberg2012} show clearly that the
nuclei $^{25,26}$O are unbound, thus making $^{24}$O the most
neutron-rich bound isotope of oxygen.  The spectroscopy of the drip
line nucleus $^{24}$O was studied in a recent
experiment~\cite{hoffman2011}. One of the exciting results of this
study is a state with an unknown spin and parity at about 7.5~MeV of
excitation energy.  Theoretical studies in this region of the nuclear
chart are challenging~\cite{volya2005,hagen2009a,otsuka2010,jdholt2010}.  Volya
and Zelevinsky~\cite{volya2005} employed an empirical two-body
shell-model interaction (above the core of $^{16}$O) and included the
particle continuum in their calculation of neutron-rich oxygen
isotopes.  Otsuka {\it et al.}~\cite{otsuka2010} included
three-nucleon forces (3NFs) within the $sd$-shell model (keeping
$^{16}$O as a core with empirical single-particle energies) and found
that three-body forces yield $^{24}$O at the neutron drip line. The
{\it ab initio} computations of neutron-rich oxygen isotopes by
Hagen~{\it et al.}~\cite{hagen2009a} employed microscopic interactions
from chiral effective field theory~\cite{entem2003}, had no core, but
were limited to nucleon-nucleon ($NN$) interactions.  Thus, we are
still lacking a complete computation of neutron rich oxygens that
properly accounts for (i) the effects of three-nucleon forces, (ii)
the presence of open decay channels and particle continuum, and (iii)
many-nucleon correlations. It is the purpose of this Letter to fill
this gap.  In particular we predict the spins and lifetimes of several
resonances in $^{24}$O at excitation energies around 7~MeV and thereby
shed light on the recent experiment~\cite{hoffman2011}.

Effective field theory (EFT) is the framework that allows for a
consistent formulation of low-energy nuclear Hamiltonians and
currents~\cite{epelbaum2009,machleidt2011,entem2003}. Within chiral
EFT, 3NFs are important contributions that enter at
next-to-next-to-leading order in the power counting. In light nuclei,
our understanding of 3NFs has improved considerably over the past
decade, and nuclear binding energies, spectra and decays cannot be
understood without
them~\cite{wiringa2002,navratil2002,navratil2009,epelbaum2011,maris2011}.
The study of the role of 3NFs in medium-mass nuclei and exotic,
neutron-rich nuclei is a frontier in contemporary nuclear structure
theory. To date, the full inclusion of 3NFs is limited to $p$-shell
nuclei. For heavier nuclei or nuclear matter, several
approaches~\cite{otsuka2010,jdholt2010,holt2010,hebeler2011,roth2011} employ a
normal-ordered approximation~\cite{hagen2007}, resulting in a
medium-dependent two-body potential that includes effects of the 3NFs.
Furthermore, the employed interactions are renormalization group
transformations~\cite{vlowk} of interactions from chiral EFT.

{\it Hamiltonian and model space.} -- The intrinsic $A$-nucleon
Hamiltonian used in this work reads
\begin{equation}
\label{ham}
\hat{H} = \sum_{1\le i<j\le A}\left({(\vec{p}_i-\vec{p}_j)^2\over 2mA} + \hat{V}_{NN}^{(i,j)} +  \hat{V}_{\rm 3N eff}^{(i,j)}\right) \,.
\end{equation}
Here, the intrinsic kinetic energy depends on the mass number $A$.
The potential $\hat{V}_{NN}$ is the chiral $NN$ interaction
developed by Entem and Machleidt \cite{entem2003} at
next-to-next-to-next-to-leading order (N$^3$LO) within chiral EFT.
The potential $\hat{V}_{\rm 3N eff}$ is the in-medium $NN$ interaction
derived by Holt~{\it et al.}~\cite{holt2010} from the leading order
chiral 3NF by integrating one nucleon over the Fermi sea (i.e. up to
the Fermi momentum $k_F$) in symmetric nuclear matter.  The leading
chiral 3NF depends on five low-energy constants (LECs).  The LECs
$c_1=-0.81$ GeV$^{-1}$, $c_3=-3.20$ GeV$^{-1}$, and $c_4=5.40$
GeV$^{-1}$ appear also in the two-pion exchange part of the chiral
$NN$ interaction and have the same values as in the N$^3$LO $NN$
potential we employ~\cite{entem2003}. The remaining LECs of the 3NF
are set at $c_D=-0.2$ and $c_E= 0.71$ together with $\Lambda_\chi=0.7$
GeV.  For the oxygen isotopes considered in this work we apply the
Fermi momentum $k_F=1.05$~fm$^{-1}$ in our potential $\hat{V}_{\rm 3N
  eff}$.  Consistent with the $NN$ force, the effective cutoff for the
3NF is $\Lambda = 500$ MeV.

Let us comment on our phenomenological two-body potential
$\hat{V}_{\rm 3N eff}$ that contains effects of 3NFs.  The
normal-ordered approximation of
3NFs~\cite{hagen2007,otsuka2010,roth2011} still requires one to
compute an enormous number of three-body matrix elements. This poses a
great challenge for the large model spaces we need to consider. The
approach of this Letter is thus simpler: The summation over the third
particle is performed in momentum space {\it before} the
transformation to the oscillator basis takes place~\footnote{For the
  approximation of two-body currents as medium-dependent one-body
  currents, this approach has been used in Ref.~\cite{menendez2011}.}.
This procedure avoids the costly computation of three-body matrix
elements in large oscillator spaces, but it introduces an uncontrolled
approximation by replacing the mean-field of a finite nucleus by that
of symmetric nuclear matter. To correct for this approximation, we
adjusted the LEC $c_E$ away from the optimal value established in
light nuclei~\cite{gazit}.

The coupled-cluster method is essentially a similarity transformation
of the Hamiltonian with respect to a reference state. This method is
accurate and efficient for nuclei with closed
(sub-)shells~\cite{ccm,dean04,bartlett07}. We compute the ground
states of $^{16,22,24,28}$O within the singles and doubles
approximation, while three-particle--three-hole (3$p$-3$h$)
excitations are included in the $\Lambda$-CCSD(T) approach of
Ref.~\cite{taube2008}.  For excited states in these closed-shell
isotopes we employ the equation-of-motion (EOM) coupled-cluster method
with singles and doubles. The open-shell nuclei $^{15,17,21,23,25}$O
are computed within the particle attached/removed EOM formalism, and
we employ the two-particle attached EOM formalism \cite{jansen2011}
for the nuclei $^{18,26}$O. For details about our implementation see
Ref.~\cite{hagen2010a}.  These EOM methods work very well for states
with dominant 1$p$-1$h$, 1$p$, 1$h$, and 2$p$ structure, respectively.
We use a Hartree-Fock basis built in 17 major oscillator shells and
varied the oscillator spacing $\hbar\omega$ between 24~MeV and 32~MeV.
Well converged energy minima are found at $\hbar\omega\approx28$~MeV
for all oxygen isotopes. Open decay channels and the particle
continuum near the dripline nucleus $^{24}$O are included within the
Gamow shell model~\cite{idbetan,michel}.  The single-particle bound
and scattering states result from diagonalizing a spherical
Woods-Saxon Hamiltonian in a discrete momentum basis in the complex
plane~\cite{michel,jensen2011}.  In the case of computing resonances
in $^{24}$O we used 35 mesh points for the $d_{3/2}$ partial wave on a
rotated/translated contour in the complex momentum plane as described
in Ref.\cite{hagen07b}.  The excited states we compute in $^{22,24}$O
are dominated by 1$p$-1$h$ excitations and continuum mixing from other
partial waves is small.  They result as solutions of a
complex-symmetric eigenvalue problem, and the imaginary part of the
energy yields the width of the state.  In computing radii we
discretized the real momentum axis with 40 points for the neutron and
proton partial waves closest to the threshold. This guarantees the
correct exponential decay of matter and charge densities at large
distances.

\begin{figure}[thbp]
  \begin{center}
    \includegraphics[width=0.45\textwidth,clip=]{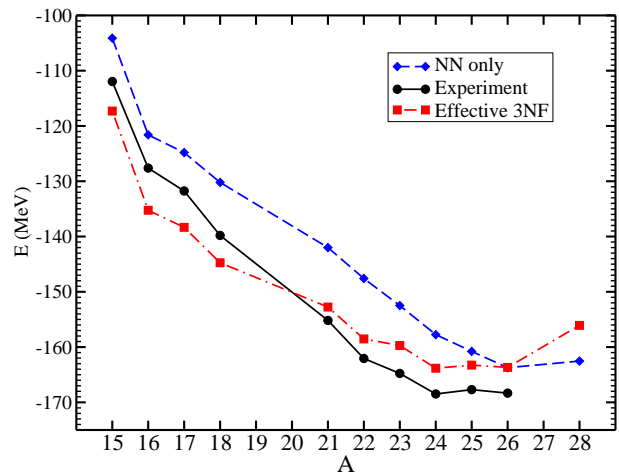}
  \end{center}
  \caption{(Color online) Ground-state energy of the oxygen isotope
    $^A$O as a function of the mass number $A$. Black circles:
    experimental data; blue diamonds: results from nucleon-nucleon
    interactions; red squares: results including the effects of
    three-nucleon forces.}
  \label{fig1}
\end{figure}

{\it Results.} -- Figure~\ref{fig1} shows the ground-state energies of
the computed oxygen isotopes (red squares) compared with experimental
data (black circles) and results limited to chiral $NN$ interactions
only (blue diamonds). For the isotopes around $^{16}$O, $NN$
interactions alone already describe separation energies rather well,
and the inclusion of effects of 3NFs mainly changes underbinding into
overbinding. For the more neutron-rich oxygen isotopes, the 3NFs
significantly change the systematics of the binding energies, and
energy differences are particularly well reproduced.  The nuclei
$^{25,26}$O are unbound with respect to $^{24}$O by about 0.4~MeV and
about 0.1~MeV, respectively, in good agreement with
experiments~\cite{hoffman2009b,lundeberg2012}.  We predict $^{28}$O to
be unbound with respect to $^{24}$O by about 4~MeV and with a resonant
width of about 1~MeV.  The extremely short life time of $^{28}$O poses
a challenge for experimental observation.  The energy difference
between light and heavy oxygen isotopes is not correctly reproduced
when compared to data. We believe that this is due to the fact that
our interaction $\hat{V}_{\rm 3N eff}$ is based on symmetric nuclear
matter. For smaller values of $k_F$, the ground-state energy of the
lighter oxygen isotopes is increased (and can be brought to good
agreement with data), while the heavier isotopes are significantly
underbound.  The value we chose for $k_F$ is thus a compromise.
   
Let us comment on our computation of oxygen isotopes with open shells.
First, we solve the CCSD equations for the Hamiltonian~(\ref{ham}) of
the closed-shell reference state, but employ the mass number $A\pm 1$
in the intrinsic kinetic energy. In a second step, we add (remove) a
neutron within the particle attached (removed) EOM.  This procedure
ensures that the final result is obtained for the intrinsic (i.e.,
translationally invariant) Hamiltonian of $^{A\pm1}$O.  The
$J^\pi=1/2^+$ ground state energy of $^{23}$O, shown in
Fig.~\ref{fig1}, resulted from using particle-removal EOM from
$^{24}$O.  For $^{18,26}$O, we performed a two-neutron attached EOM
computation based on the reference states for $^{16,24}$O, the latter
being computed with the mass number $A=18,26$ in the intrinsic kinetic
energy.  This approach is unproblematic for separation energies but it
introduces an error in the computation of resonance widths. Our
computation of $^{25}$O within the neutron attached EOM employs a
Gamow basis. Here the continuum threshold is incorrectly set by the
closed-shell reference of $^{24}$O, computed with the mass number
$A=25$ in the intrinsic kinetic energy. Clearly, this introduces a
small error by shifting the scattering threshold, and thereby affects
the widths of resonance states that are very close to the threshold.
In Ref.~\cite{hagen09} we showed that the coupled-cluster wave
function factorizes into an intrinsic and a center-of-mass part. The
center-of-mass wave function is to a very good approximation a
Gaussian with a frequency $\hbar\tilde\omega \approx 14$~MeV for
$^{24}$O. Thus, we estimate the error introduced in the scattering
threshold of $^{25}$O to be ${3\over
  4}\hbar\tilde\omega/A\approx$~0.4~MeV.

\begin{figure}[thbp]
  \begin{center}
    \includegraphics[width=0.45\textwidth,clip=]{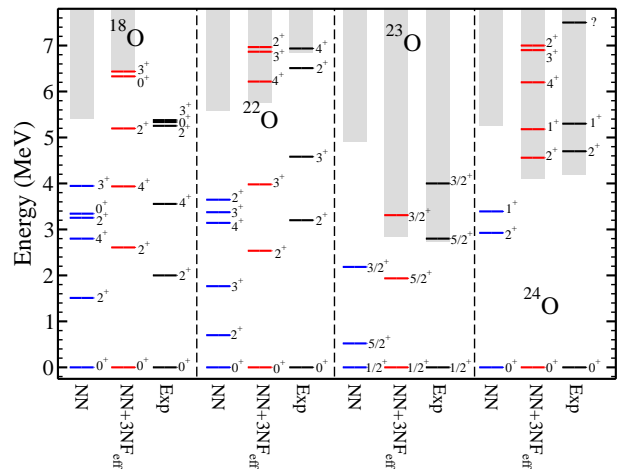}
  \end{center}
  \caption{(Color online) Excitation spectra of oxygen isotopes
    computed from chiral nucleon-nucleon interactions, with inclusion
    of the effects of three-nucleon forces, and compared to data.}
  \label{fig2}
\end{figure}

Figure~\ref{fig2} shows the excitation spectra of neutron-rich oxygen
isotopes and compares the results limited to chiral $NN$ interactions
(blue lines) to the results obtained with our inclusion of 3NFs (red
lines), and to experimental data (black lines)~\footnote{The displayed
  experimental levels are dominated by 1$p(h)$, 2$p$ or 1$p$-1$h$
  excitations.}.  The particle continua above the scattering
thresholds are shown as gray bands. Calculations limited to $NN$
interactions yield the correct level ordering but very compressed
spectra when compared to data, and all the computed excited states are
well bound with respect to the neutron emission thresholds.  However,
the inclusion of 3NFs increases the level spacing and significantly
improves the agreement with experiment.  Several of the excited states
are resonances in the continuum, and the proximity of the continuum is
particularly relevant for the dripline nucleus $^{24}$O. Here, the
Gamow basis is essential for a proper description of the excited
resonant states.  In $^{24}$O we find three resonant states near the
unknown experimental state at about 7.5~MeV of excitation
energy~\cite{hoffman2011}.  The excited $J^\pi=3/2^+$ state in
$^{23}$O is computed as a neutron attached to $^{22}$O, while the
excited $J^\pi=5/2^+$state is computed by neutron removal from
$^{24}$O. Since we are interested in the excitation energy relative to
the ground state we compute the $J^\pi=1/2^+$ ground state either by
adding or removing a particle, consistent with the particular excited
state.  For the lighter isotopes $^{15,16,17}$O, our inclusion of 3NFs
yields only smaller changes to the spectra when compared with results
from $NN$ interactions only.

For the closed-shell nucleus $^{24}$O we also computed resonance
widths of excited states that are dominated by 1$p$-1$h$ excitations
from the $s_{1/2},d_{5/2}$ hole to the $d_{3/2}$ particle orbitals.
Table~\ref{tab2} shows that the first $J^\pi=2^+$ and $1^+$ excited
states agree well with experimental data, both for the excitation
energy and the resonance widths.  The rather small widths and
quasi-bound nature of these states can be attributed to the large
angular momentum barrier of the $d_{3/2}$ orbital, together with
neutron pairing effects. Above these $2_1^+$ and $1_1^+$ resonances we
find several states with spin and parity $J^\pi = 1^+$ to $4^+$ and
excitation energies ranging from $6.2$~MeV to $8.4$~MeV.  The small
ratio $E_{4_1^+}/E_{2_1^+} \approx 1.36$ and the relatively high
energy $E_{2_1^+}$ lend theoretical support to the doubly magic nature
of $^{24}$O~\cite{kanungo2009}.  The low experimental resolution of
the resonance at 7.5~MeV let Hoffman {\it et al.}~\cite{hoffman2011}
to speculate that this resonance could be a superposition of narrow
resonances with spins and parity $J^\pi = 1^+$ to $4^+$.  Our
calculation clearly supports this suggestion, except for the $1_2^+$
state which we find at $8.4$~MeV of excitation energy and with a width
of $0.56$~MeV.

\begin{figure}[thbp]
  \begin{center}
    \includegraphics[width=0.45\textwidth,clip=]{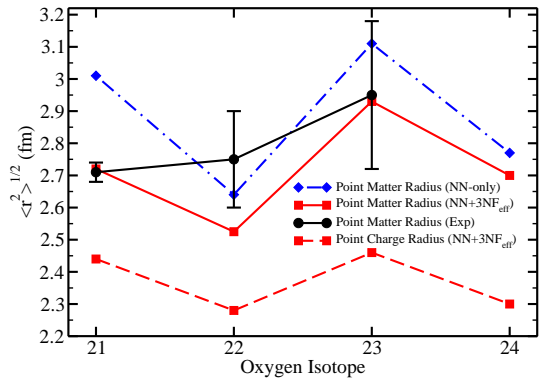}
  \end{center}
  \caption{(Color online)Point matter and point charge radii 
    of neutron rich oxygen isotopes
    computed from chiral nucleon-nucleon interactions, with inclusion
    of the effects of three-nucleon forces, and compared to data.}
  \label{fig3}
\end{figure}

\begin{table}
  \begin{ruledtabular}
  \begin{tabular}{|l||l|l|l|l|l|l|}
   $J^\pi$ & $2_1^+$ & $1_1^+$     & $4_1^+$ & $3_1^+$ & $2_2^+$ & $1^+_2$ \\ \hline
    $E_{\rm CC}$    & 4.56           &  5.2          & 6.2       & 6.9  & 7.0 &  8.4 \\ 
    $E_{\rm Exp}$   & 4.7(1)  &  5.33(10)&           &    &   &   \\ \hline
    $\Gamma_{\rm CC}$ & 0.03 & 0.04 & 0.005 & 0.01  & 0.04 &  0.56  \\
    $\Gamma_{\rm Exp}$ & $0.05^{+0.21}_{-0.05}$&
    $0.03^{+0.12}_{-0.03}$ & & & &  \\
  \end{tabular}
  \caption{Excited states in $^{24}$O computed within EOM-CCSD
    compared to experimental data from Ref.~\cite{hoffman2011}. Energies and widths are
    in MeV.}
  \label{tab2}
  \end{ruledtabular}
\end{table} 
Figure~\ref{fig3} shows the computed point matter and point charge
radii for the neutron-rich isotopes $^{21-24}$O with a comparison to
the experimental data (Ref.~\cite{ozawa01} for $^{21}$O and
Ref.~\cite{kanungo2011} for $^{22,24}$O). This computation employs the
intrinsic density with respect to the center of mass. 
Our computed radii agree very well with experiment for the odd
isotopes $^{21,23}$O, while for $^{22}$O we underestimate the radii
compared to experiment. We also computed the point matter radii from
$NN$ interactions only (blue diamonds). In this case the radii
overestimate the data for $^{21,23}$O, while $^{22,24}$O they are
closer to the results with effects of 3NF's included. The computed
charge radii clearly exhibit an odd-even staggering consistent with
the shell closures at neutron numbers $N=14,16$.  For $^{16}$O, the
computed point matter and charge radii are $2.23$~fm and $2.24$~fm,
respectively. This is about $0.3$~fm smaller than experiment and
consistent with the computed overbinding and the increased neutron and
proton separation energies.

{\it Summary.} -- We employed interactions from chiral effective field
theory, performed coupled-cluster computations of oxygen isotopes, and
included effects of the particle continuum and of three-nucleon
forces.  Three-nucleon forces were approximated as in-medium
nucleon-nucleon forces. This approach is computationally feasible and
in keeping with the spirit of effective field theory.  Compared to
computations based on nucleon-nucleon interactions alone, the included
3NFs yield a significant improvement in binding energies and spectra.
Our results confirm that chiral interactions yield the neutron drip
line at $^{24}$O, and we are able to compute spin, parities and
resonance widths for several excited states close to the dripline. In
particular, we compute several long-lived resonances at about 7~MeV of
excitation energy in $^{24}$O.

\begin{acknowledgments}
  This work was supported by the Office of Nuclear Physics,
  U.S.~Department of Energy (Oak Ridge National Laboratory), and 
  through the Grants Nos.~DE-FG02-03ER41270 (University of Idaho),
  DE-FG02-96ER40963 (University of Tennessee), and DE-FC02-07ER41457
  (UNEDF SciDAC).  This research used computational resources of the
  National Center for Computational Sciences, the National Institute
  for Computational Sciences, and the Notur project in Norway.
\end{acknowledgments}

\end{document}